\documentclass[10pt,twocolumn]{article}
\pdfoutput=1
\usepackage{cite}
\usepackage{amsmath,bbm}
\usepackage{amssymb}
\usepackage[latin1]{inputenc}
\usepackage{mathrsfs}
\usepackage{mathtools}
\usepackage[dvipsnames]{xcolor}
\usepackage[footnotesize]{caption}
\usepackage{xcolor}
\usepackage{dsfont}
\usepackage{hyperref}
\usepackage[hmarginratio=1:1,top=32mm,columnsep=20pt]{geometry}
\usepackage{paralist}
\usepackage{abstract}
\usepackage{upgreek}
\usepackage{soul}

    \makeatletter
    \let\@fnsymbol\@arabic
    \makeatother

\newcommand{\bea}{\begin{eqnarray}}
\newcommand{\eea}{\end{eqnarray}}
\newcommand{\be}{\begin{equation}}
\newcommand{\ee}{\end{equation}}
\newcommand{\ba}{\begin{array}}
\newcommand{\ea}{\end{array}}

\def\gsim{\mathrel{\rlap{\lower4pt\hbox{\hskip1pt$\sim$}}
    \raise1pt\hbox{$>$}}}

\title{\vspace{-15mm}
	\fontsize{16pt}{10pt}\selectfont
	\textbf{Resolvable heavy neutrino-antineutrino \\[1mm]  oscillations at colliders}
	}	
\author{%
	\large
	\textsc{Stefan~Antusch$^{\star \dagger}$
	,  Eros~Cazzato$^\star$
	, Oliver~Fischer$^{\star\text{\textdaggerdbl}}$
	}\\[10pt]
	\normalsize	$^\star$ Department of Physics, University of Basel, \\ 
	\normalsize 	Klingelbergstr.\ 82, CH-4056 Basel, Switzerland\\[5pt]
	\normalsize	$^\dagger$ Max-Planck-Institut f\"ur Physik (Werner-Heisenberg-Institut),\\
	\normalsize	F\"ohringer Ring 6, D-80805 M\"unchen, Germany\\[5pt]
	\normalsize 	$^\text{\textdaggerdbl}$ Institute for Nuclear Physics, Karlsruhe Institute of Technology, \\
	\normalsize	Hermann-von-Helmholtz-Platz 1, D-76344 Eggenstein-Leopoldshafen, Germany
	\vspace{-5mm}
	}
\date{}

\begin{document}

\maketitle

\begin{abstract}
\noindent 
Heavy neutrino-antineutrino oscillations can appear in mechanisms of low scale neutrino mass generation, where pairs of heavy neutrinos have almost degenerate masses. We discuss the case where the heavy neutrinos are sufficiently long-lived to decay displaced from the primary vertex, such that the oscillations of the heavy neutrinos into antineutrinos can potentially be observed at the (high-luminosity) LHC and at currently planned future collider experiments. The observation of these oscillations would allow to measure the mass splitting of the respective heavy neutrino pair, providing a deep insight into the nature of the neutrino mass generation mechanism.

\end{abstract}

\noindent
\paragraph{Introduction:}

Sterile neutrinos, which are singlets under the gauge group of the Standard Model (SM), are an attractive extension of the SM to explain the observed masses of the light neutrinos. We consider scenarios where the masses of the heavy neutrinos are around the electroweak (EW) scale. Various models with such a low scale seesaw mechanism exist in the literature (see e.g.\ \cite{Wyler:1982dd,Mohapatra:1986bd,Malinsky:2005bi,Shaposhnikov:2006nn,Kersten:2007vk,Gavela:2009cd,Chattopadhyay:2017zvs}). 

The models may be classified by how the approximate ``lepton number''-like symmetry, which ensures the smallness of the light neutrinos' masses, is broken: If the breaking happens in the sterile neutrino mass matrix, the models are called low scale ``inverse seesaw'' models \cite{Wyler:1982dd,Mohapatra:1986bd}, and when the breaking happens in the coupling matrix between the sterile neutrinos and the active SM neutrinos, i.e.\ in the Yukawa sector, then the models are called low scale ``linear seesaw'' models \cite{Malinsky:2005bi}. In both classes of models, due to the approximate ``lepton number''-like symmetry, pairs of heavy neutrinos have almost degenerate masses, i.e.\ they form heavy pseudo-Dirac particles. 

For collider studies it is often useful to focus on one of the pairs of heavy neutrinos and to consider the limit of intact ``lepton number''-like symmetry, as e.g.\ in the ``SPSS'' (Symmetry Protected Seesaw Scenario) benchmark model (cf.\ \cite{Antusch:2015mia}). A recent study of the prospects for testing such low scale seesaw models at the LHC and at future colliders can be found e.g. in \cite{Antusch:2016ejd}. 

The small mass splittings between the heavy neutrinos typical for low scale seesaw models lead to oscillations between the heavy neutrinos and their antiparticles. The time-integrated effect of these oscillations, i.e.\ when the oscillations cannot be resolved experimentally, has been highlighted in \cite{Gluza:2015goa,Dev:2015pga,Anamiati:2016uxp,Das:2017hmg}. 
Heavy sterile neutrino oscillations in meson decays have been discussed in \cite{Boyanovsky:2014una,Cvetic:2015ura}, and further observable effects at colliders e.g.\ in \cite{Cvetic:2018elt}.
In this letter, we discuss the case where the heavy neutrinos are sufficiently long-lived to decay displaced from the primary vertex. We show that the oscillations of the heavy long-lived neutrinos can potentially be resolved at the high luminosity (HL) phase of the LHC and at currently planned future collider experiments. The observation of these oscillations would allow to measure the mass splitting of the respective heavy neutrino pair, providing a deep insight into the nature of the neutrino mass generation mechanism. Such a measurement would also be very important towards testing leptogenesis in minimal seesaw models, as discussed recently in \cite{Drewes:2016gmt,Drewes:2016jae,Antusch:2017pkq}.

\paragraph{Low scale seesaw mechanism:}
After electroweak symmetry breaking the Lagrangian density containing the masses of the light and heavy neutrinos can be written as $\mathscr{L}_{\rm mass} = -1/2 \overline{\Psi^c} M_\nu \Psi + H.c.$, where the active neutrinos $\nu_L$ and pairs of sterile neutrinos $(N_R^1)^c,(N_R^2)^c$ have been combined to $\Psi=\left(\nu_L , (N_R^1)^c , (N_R^2)^c\right)^T$. If the approximate ``lepton number''-like symmetry mentioned above was exact, the neutrino mass matrix is schematically given by
\begin{equation}
M_\nu = \begin{pmatrix}
0 & m_D & 0 \\
(m_D)^T & 0 & M  \\
0 & M & 0
\end{pmatrix}.  \label{eq:Mnu}
\end{equation}
In the following, we will consider one pair of sterile neutrinos for simplicity. In the exact symmetry limit it has a Dirac-type mass $M$,\footnote{In the general case with more sterile neutrinos, $M$ stands for a $n \times n$ Dirac-type mass matrix of $n$ pairs of sterile neutrinos.} and $m_D = Y_\nu v_\mathrm{EW}/\sqrt{2}$ with the neutrino Yukawa coupling matrix $Y_\nu$ coupling the three active neutrinos to the sterile neutrino which carries the same charge under the ``lepton number''-like symmetry.

In order to generate the light neutrino masses, the structure in eq.~(\ref{eq:Mnu}) has to be perturbed. 
There are two characteristic ways to implement such perturbations, corresponding to two classes of models:
\begin{itemize}

\item {\bf Linear Seesaw:} The perturbation is introduced in the 1-3 (and 3-1) block of the symmetric matrix $M_\nu$ in eq.~(\ref{eq:Mnu}):
\begin{equation}
M^\mathrm{lin}_\nu = \begin{pmatrix}
0 & m_D & m^\prime_D \\
(m_D)^T & 0 & M  \\
(m^\prime_D)^T & M & 0
\end{pmatrix},  \label{eq:MnuLin}
\end{equation}
where $m_D^\prime$ has small entries compared to $m_D$. 

With two sterile neutrinos (one pair), the symmetry breaking term $m_D^\prime$ can generate masses for two of the three light neutrinos as well as a splitting $\Delta M^\mathrm{lin}$ between the masses of the two heavy neutrinos. With an appropriate choice of $m_D$ and $m_D^\prime$, the present neutrino oscillation data can be accommodated. We will refer to this scenario as the ``minimal linear seesaw model''.

It can be shown that in the minimal linear seesaw scenario the mass splitting $\Delta M^\mathrm{lin}$ is predicted in terms of the measured mass squared differences of the light neutrinos, which yields, in the case of normal mass ordering (NO) or inverse mass ordering (IO):
\begin{align}
&\Delta M^\mathrm{lin}_{\rm NO} \!=\!\! \tfrac{2\rho_\mathrm{NO}\sqrt{\Delta m^2_{21}}}{1-\rho_\mathrm{NO}} = \Delta m_{32}= 0.0416\text{ eV} \label{eq:DeltaMlinNO} \\
&\Delta M^\mathrm{lin}_{\rm IO} \!=\!\! \tfrac{2\rho_\mathrm{IO}\sqrt{\Delta m^2_{23}}}{1+\rho_\mathrm{IO}}=\Delta m_{21} \! =\! 0.000753\text{ eV}\label{eq:DeltaMlinIO} 
\end{align}
with $ \rho_\mathrm{NO}  =\frac{\sqrt{r+1}-\sqrt{r}}{\sqrt{r}+\sqrt{r+1}}$ and $r=\frac{\left|\text{$\Delta $m}_{21}^2\right|}{\left|\text{$\Delta $m}_{32}^2\right|}$, and with $\rho_\mathrm{IO}  =\frac{\sqrt{r+1}-1}{\sqrt{r+1}+1}$ and  $r=\frac{\left|\text{$\Delta $m}_{21}^2\right|}{\left|\text{$\Delta $m}_{13}^2\right|}$, where $\Delta m^2_{ij}$ corresponds to the mass squared differences $m^2_{\nu_{i}}-m^2_{\nu_{j}}$ and $\Delta m_{ij}$ to the mass splitting $m_{\nu_{i}}-m_{\nu_{j}}$, with $m_{\nu_{i}}$ labelling the light neutrino masses. In the last step, the mass of the lightest of the light neutrino has been set to zero, as implied by the minimal linear seesaw model.

\item {\bf Inverse Seesaw:} The perturbation is introduced in the 3-3 block of the symmetric matrix $M_\nu$ in eq.~(\ref{eq:Mnu}):
\begin{equation}
M^\mathrm{inv}_\nu = \begin{pmatrix}
0 & m_D & 0 \\
(m_D)^T & 0 & M  \\
0 & M & \mu
\end{pmatrix},  \label{eq:MnuInv}
\end{equation}
where $\mu \ll M$ violates lepton number and introduces a mass for {\em one} of the light neutrinos (which may for instance be the lightest or the heaviest of the light neutrinos) as well as a mass splitting between the two heavy neutrinos. We emphasise that in contrast to the minimal linear seesaw model discussed above, additional sterile neutrinos (e.g.\ additional pairs) have to be introduced in order to be consistent with neutrino oscillation results. To obtain an estimate for the sterile neutrino mass splitting in the inverse seesaw case, we will assume that the considered pair of sterile neutrinos dominantly generates one of the masses of the light neutrinos, labeled $m_{\nu_{i}}$.

Since for the case of two sterile neutrinos considered here one can choose $m_D$, $M$ and $\mu$ real and positive without loss of generality, the mass $m_{\nu_{i}}$ is (to leading order) related to the squared sum $|\theta|^2 := (m_D^\dagger m_D^{})/M^2$
of the active-sterile mixing angles by 
\begin{equation}
m_{\nu_{i}} = \mbox{Tr} \left(\mu \frac{m_D^{} m_D^T}{M^2}\right) = \mu \frac{m_D^\dagger m_D^{}}{M^2} 
= \mu |\theta|^2 \:,
  \label{eq:mnui}
\end{equation}
The mass splitting $\Delta M^\mathrm{inv}$ then satisfies  
\begin{equation} 
\Delta M^\mathrm{inv} = \mu - m_{\nu_i} \approx \frac{m_{\nu_{i}}}{ |\theta|^2} \:.
  \label{eq:MnuInvDelta}
\end{equation}
In the following we will use this relation with $m_{\nu_{i}}$ in the range $m_{\nu_{i}} =0.1\:\mbox{eV} \dots 10^{-4}\:\mbox{eV}$ to give example values for $\Delta M^\mathrm{inv}$ as a function of $|\theta|^2$. We like to emphasise that the resulting ranges for $\Delta M^\mathrm{inv}$ are no strict predictions but at best guidelines for the inverse seesaw scenario. A smaller mass splitting could be a consequence of an even smaller  $m_{\nu_{i}}$, and a larger mass splitting could be the result of cancellations between the contributions of two pairs of sterile neutrinos to the light neutrino mass matrix. \footnote{Furthermore, we note that in addition a perturbation of $M_\nu$ of eq.~(\ref{eq:Mnu}) in the 2-2 block may be introduced which does not lead to a contribution to the light neutrinos' masses (at tree-level) but can enhance or reduce the splitting $\Delta M^\mathrm{inv}$.}

\end{itemize}

In summary, we have obtained estimates for the typical mass splittings $\Delta M$ of the almost degenerate sterile neutrinos in low scale seesaw scenarios as functions of the light neutrino masses (respectively the mass splittings). In case of the minimal linear seesaw model, the values $\Delta M^\mathrm{lin}_{\rm NO}$ and $\Delta M^\mathrm{lin}_{\rm IO}$ are predictions, whereas in the inverse seesaw case or the general linear seesaw case with more pairs of sterile neutrinos, one should view the given values as guidelines only.

\paragraph{Heavy neutrino-antineutrino oscillations:} When heavy neutrinos are produced from $W$ decays together with charged leptons or antileptons, we refer to them as heavy antineutrinos $\overline N$ or neutrinos $N$, respectively. When they decay via the charged current, they again produce either a lepton or an antilepton, $N \to \ell^- W^+$ or $\overline N \to \ell^+ W^-$. 

If the ``lepton number''-like symmetry is intact, i.e.~without its breaking to give light neutrinos its mass, processes with the heavy neutrinos at colliders are lepton number conserving (LNC). For instance at proton-proton ($pp$) colliders, there would be only LNC processes $pp \to \ell^+_\alpha \ell^-_\beta j j$ but no lepton number violating (LNV) processes $pp \to \ell^\pm_\alpha \ell^\pm_\beta j j$. We will focus on these processes as an example in the following, since they can yield an unambiguous signal of LNV at $pp$ colliders. 

In the presence of LNV perturbations  in the mass matrix of eq.~(\ref{eq:Mnu}) however, also LNV processes $pp \to \ell^\pm_\alpha \ell^\pm_\beta j j$ are possible. One can view these events as stemming from $N$ (or $\overline N$) being produced together with a charged antilepton (or lepton) which then oscillates into a $\overline N$ (or $N$), decaying into a charged antilepton (or lepton), finally producing a lepton-number violating final state.

\begin{figure}
\centering
\includegraphics[width=0.4\textwidth]{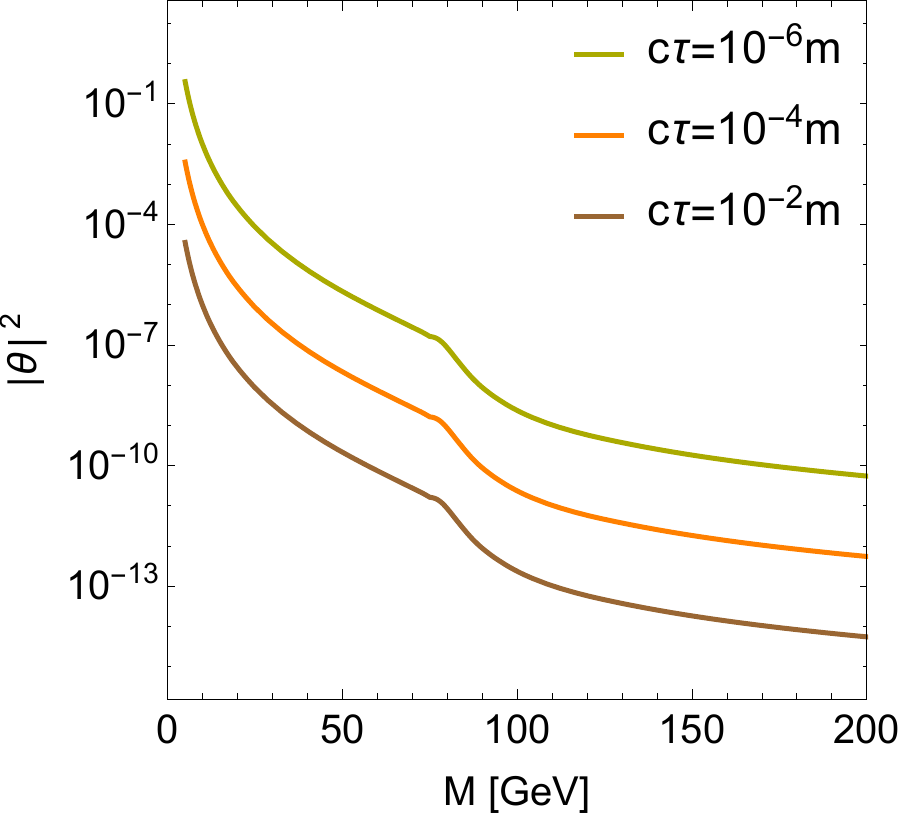}
\caption{\label{fig:lifetime}
Contours of constant decay length of the heavy neutrinos $x = \tau \,c$ in the proper frame, where $\tau$ is the lifetime in the proper frame (cf.\ discussion in section 3 of \cite{Antusch:2016vyf}). The decay length in the laboratory frame is given by $x \sqrt{\gamma^2 -1}$ with the Lorentz factor $\gamma$.}
\end{figure}

When the heavy neutrinos have sufficiently small decay widths, they can have macroscopic lifetimes such that their decay occurs displaced from the primary vertex, which allows for powerful searches and opens up the possibility to observe the oscillation patterns in the decay spectra. We show in figure \ref{fig:lifetime} for which parameters $M$ and $|\theta |^2$ macroscopic lifetimes are possible.

Due to heavy neutrino-antineutrino oscillations, following \cite{Anamiati:2016uxp,ReviewNir}, the ratio between LNV and LNC events between times $t_1$ and $t_2$ after heavy neutrino production will be referred to as $R_{\ell \ell} (t_1,t_2)$ and is given as:
\begin{equation}
R_{\ell \ell} (t_1,t_2)= \frac{\int_{t_1}^{t_2} |g_- (t)|^2 dt}{\int_{t_1}^{t_2} |g_+ (t)|^2 dt} = \frac{\#(\ell^+ \ell^+) + \#(\ell^- \ell^-)}{\#(\ell^+ \ell^-)}\:,
  \label{eq:R_ellell}
\end{equation}
where $g_-(t)\simeq- i e^{- i M t} e^{-\frac{\Gamma}{2}t} \sin\left(\frac{\Delta M}{2} t\right)$, $g_+(t)\simeq e^{- i M t} e^{-\frac{\Gamma}{2}t} \cos\left(\frac{\Delta M}{2} t\right)$ and where $\Gamma$ is the heavy neutrino decay width. $|g_-(t)|^2$ corresponds to the time-dependent probability that a heavy neutrino has oscillated into a heavy antineutrino and vice versa, and $|g_+(t)|^2$ denotes the probability that no oscillation has occurred.\footnote{We note that here we neglect CP violating effects, which can be introduced by perturbations of the mass matrix of eq.~(\ref{eq:Mnu}) and could leave imprints in the distribution of the $\ell^\pm_\alpha \ell^\pm_\beta j j$ and  $\ell^+_\alpha \ell^-_\beta j j$ final states.}

From the above formula, we can see that the oscillation period of the heavy neutrinos is given by $t_\mathrm{osc} = \frac{4 \pi}{\Delta M}$. In the minimal linear seesaw scenario (using Eqs.~(\ref{eq:DeltaMlinNO}) and (\ref{eq:DeltaMlinIO})) and with our estimates for the inverse seesaw scenario from Eq.~(\ref{eq:MnuInvDelta}), we obtain for the oscillation length in the laboratory system:
\begin{align}
&\lambda_\mathrm{osc}^\mathrm{lin,NO} = 5.96 \cdot 10^{-5} \sqrt{\gamma^2 -1} \:\text{m}\:, \label{eq:OscLNO}\\
&\lambda_\mathrm{osc}^\mathrm{lin,IO} = 3.29 \cdot 10^{-3} \sqrt{\gamma^2 -1}\: \text{m}\:,\label{eq:OscLIO}\\
&\lambda_\mathrm{osc}^\mathrm{inv}  \approx 2.48 \cdot 10^{-6} \left(\frac{|\theta|^2}{10^{-4}}\right) \left( \frac{10^{-4}\:\text{eV}}{m_{\nu_{i}}}\right) \sqrt{\gamma^2 -1} \:\text{m}\:.
\end{align}
Especially when the Lorentz factor is large, the oscillation length in the laboratory system can be large enough to be resolved in an experiment. The case of the minimal linear seesaw with IO looks particularly promising in this context. For observability it is also important that the decay of the heavy neutrinos is sufficiently displaced from the primary vertex (cf.\ figure \ref{fig:lifetime}).

We remark that the oscillations can only occur if the mass eigenstates are in a coherent superposition, which imposes the following conditions \cite{Kayser:1981ye,Akhmedov:2007fk}: Coherence in the production and detection point requires that the quantum mechanical uncertainty in the mass squared of the heavy neutrino $\sigma_M^2$ is larger than the mass squared differences $\Delta(M^2)=2M\Delta M$. The uncertainty $\sigma_M^2$ is proportional to $E_N\,\Gamma_W$, with $E_N$ being the average energy of the heavy neutrino and $\Gamma_W$ the width of the parent particle of the heavy neutrino, which is the $W$ boson. Coherence during propagation, i.e.~preservation of coherence due to wave packet separation, is fulfilled for the maximum coherence length, which corresponds to the minimum group velocity difference, $x_{\rm coh}\simeq[\Gamma_W\,\Delta (v_g)]^{-1}$. For heavy neutrino masses, mass splittings, and boosts  in the ranges where displaced vertices from heavy neutrino decays could be detected at present and future colliders (cf.\ \cite{Antusch:2016ejd}), the coherence conditions are satisfied.

\paragraph{Potentially observable oscillations:}
As an example, we now consider the LHCb experiment and heavy neutrinos with $M=7$ GeV, $|\theta|^2=10^{-5}$. These parameter values are consistent with the present bounds, cf.\ e.g.\ \cite{Abreu:1996pa,Liventsev:2013zz,Aaij:2014aba,Shuve:2016muy}, and within reach of, e.g., the HL phase of the LHC  \cite{Antusch:2017hhu}.

\begin{figure}
\centering
\includegraphics[width=0.4\textwidth]{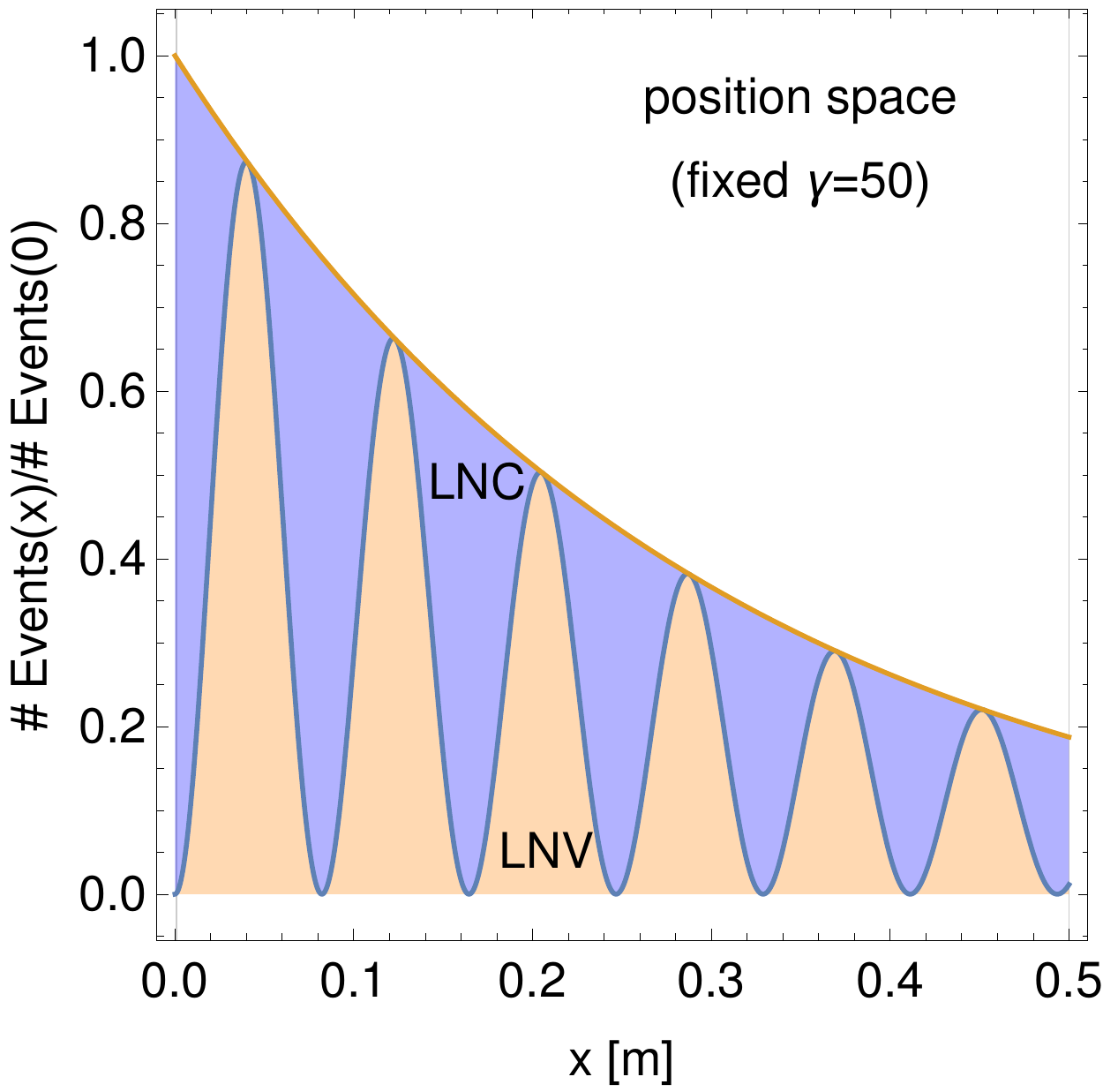}

\includegraphics[width=0.4\textwidth]{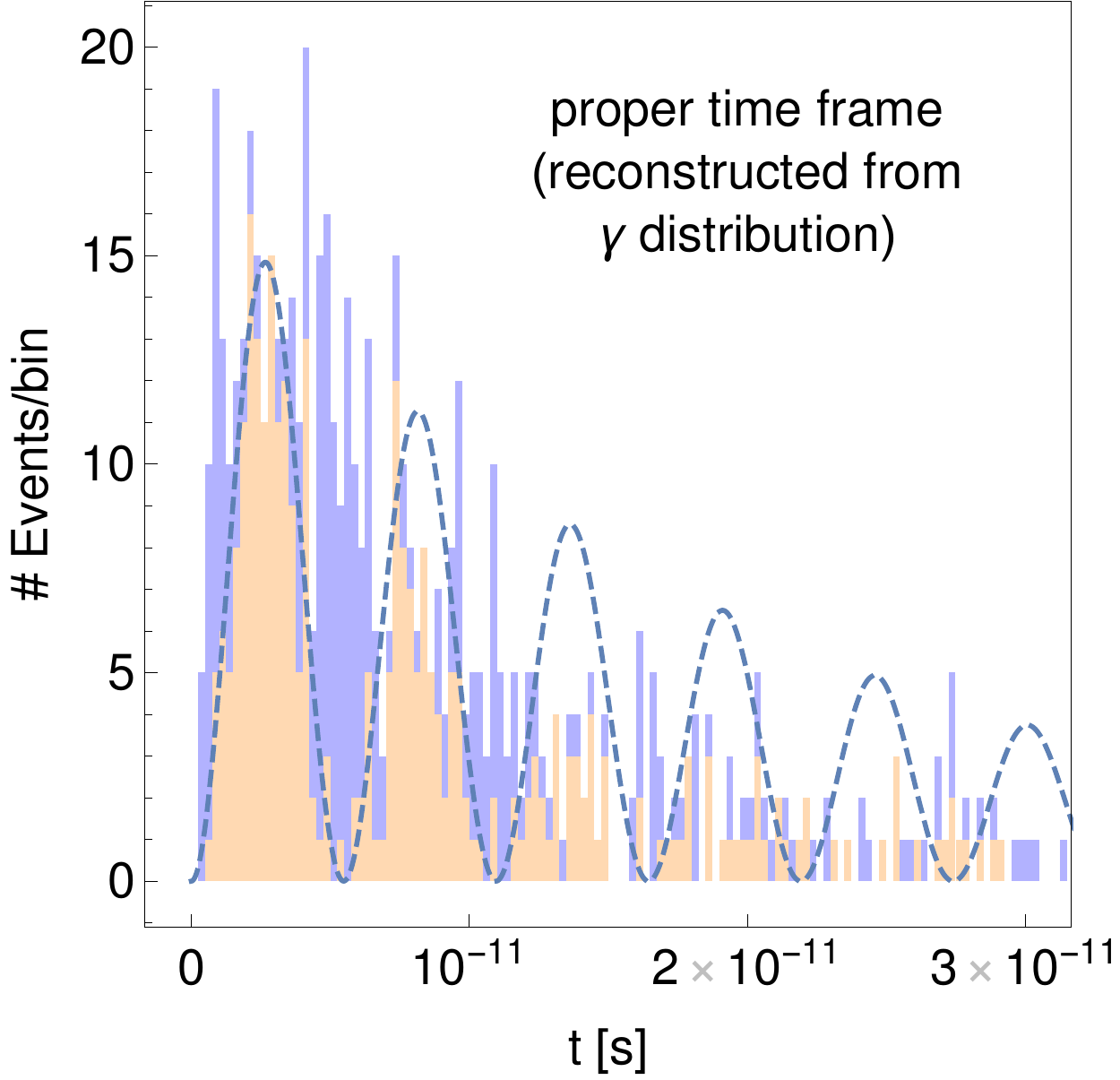}
\caption{\label{fig:OscAtLHCb} Signature of heavy neutrino-antineutrino oscillations at LHCb with $M=7$ GeV, $|\theta|^2=10^{-5}$, and minimal linear seesaw with IO for the light neutrino masses. \\
\emph{Top:} Idealized picture in the laboratory frame with an assumed fixed Lorentz factor $\gamma=50$. The orange envelope curve shows the total number of events at distance $x$ from the primary vertex, for the process $pp \to \ell \ell j j $, normalised to the number of events at $x=0$. The orange and blue areas denote the fractions of LNV and LNC events, respectively. The shown range for the x-axis, up to 50 cm, corresponds to the size of LHCb's VErtex LOcator (VELO).\\
\emph{Bottom:} Reconstructed example oscillation pattern as a function of time in the proper frame where the effect of the $\gamma$ distribution is accounted for. We created a Monte Carlo sample of 620 displaced vertex events between 2 and 50 cm, a region that is free of backgrounds for $\mu jj$ final states at LHCb, see ref.~\cite{Aaij:2016xmb} for details. This number corresponds to the number of events during the HL phase (run 5 with an assumed integrated luminosity of 380 fb$^{-1}$) when considering only muon final states. We took into account a resolution for the displacement measurement of 3 mm, and an error on the measurement of the Lorentz factor of 10\% (for a discussion cf.\ \cite{Aaij:2014jba}). We note that after the first few oscillations the pattern is smeared out due to the resolution, and the error on $\gamma$. 
}
\end{figure}

For the case of light neutrino masses from a low scale minimal linear seesaw mechanism with IO, the results for the fractions of LNV and LNC events, as function of the distance $x$ from the primary vertex, are shown as an idealized plot assuming a fixed Lorentz factor of $\gamma=50$ (which is typical for these parameters as discussed in \cite{Antusch:2017hhu}) in the top of figure \ref{fig:OscAtLHCb}.

In reality the distribution of the Lorentz factor leads to a smearing out of the oscillation pattern in position space.  
Nevertheless, the oscillation pattern as function of time in the proper frame of the heavy neutrinos can be reconstructed by measuring the $\gamma$ for each event (and using the reconstructed heavy neutrino mass scale $M$). 
An example for the estimated reconstruction in the proper frame is shown in the bottom of figure \ref{fig:OscAtLHCb} for the HL phase of LHCb. 
This illustrates that there are promising conditions to observe the signature of heavy neutrino oscillations at LHCb or at future colliders. To confirm the feasibility, a more detailed study of the detector response and its efficiencies for the relevant detector regions would be required.

We note that for the case of NO of the light neutrino masses the oscillation length is shorter (cf.\ eq.\ \eqref{eq:OscLNO}). This is in principle still resolvable with the LHCb tracking resolution, however we expect the experimental uncertainties to make this more difficult than for the IO case. 

For the inverse seesaw case with $m_{\nu_{i}}$ in the range $m_{\nu_{i}} =0.1\:\mbox{eV} \dots 10^{-4}\:\mbox{eV}$ (using our estimate from Eq.~(\ref{eq:MnuInvDelta})) and for the used example parameter point, the oscillations would not be visible since the oscillation length is estimated to be much below the tracking resolution of ${\cal O}(10^{-4})$ m \cite{Aaij:2014jba}. 
 
Regarding $pp$ colliders, of course also the FCC-hh \cite{Golling:2016gvc} and SppC \cite{CEPC-SPPCStudyGroup:2015csa} would be very promising for testing these signatures with possible much higher $\gamma$ factors and higher luminosity. Other colliders where signatures of heavy neutrino-antineutrino oscillations could be studied are the electron-proton colliders, e.g.\ the LHeC \cite{AbelleiraFernandez:2012cc} or the FCC-eh \cite{Bruening:2013bga}, since they also provide unambiguous signal processes for LNV (see e.g.\ \cite{Antusch:2016ejd}).    

\begin{figure}[h!]
\centering
\includegraphics[width=0.4\textwidth]{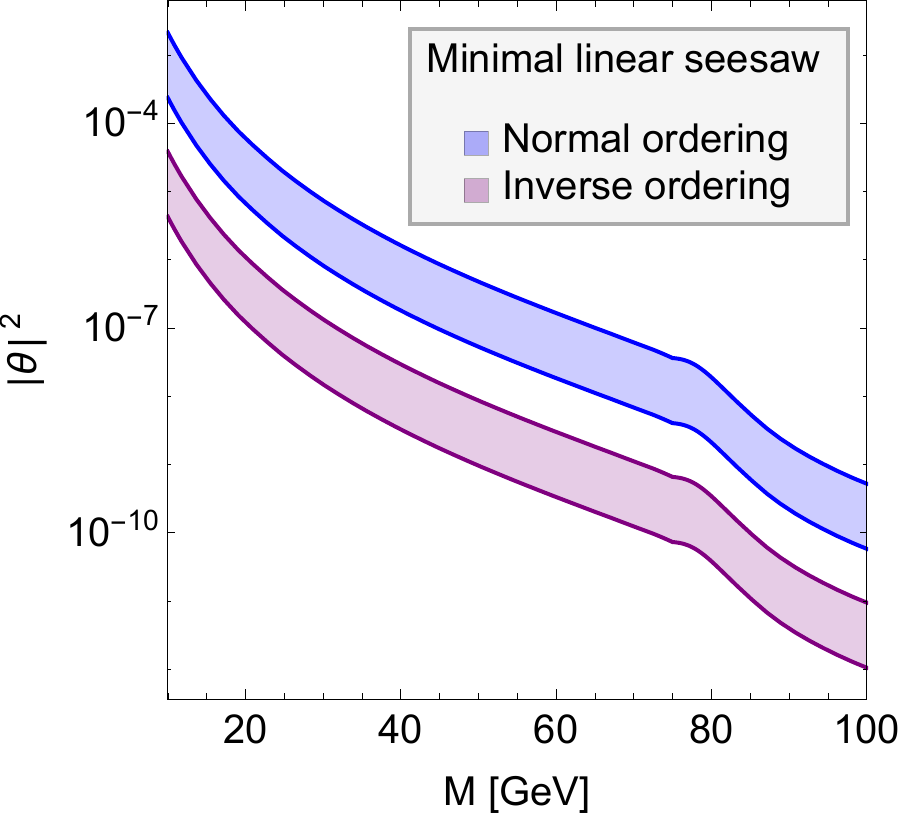}
\medskip
\vspace{4mm}
\includegraphics[width=0.4\textwidth]{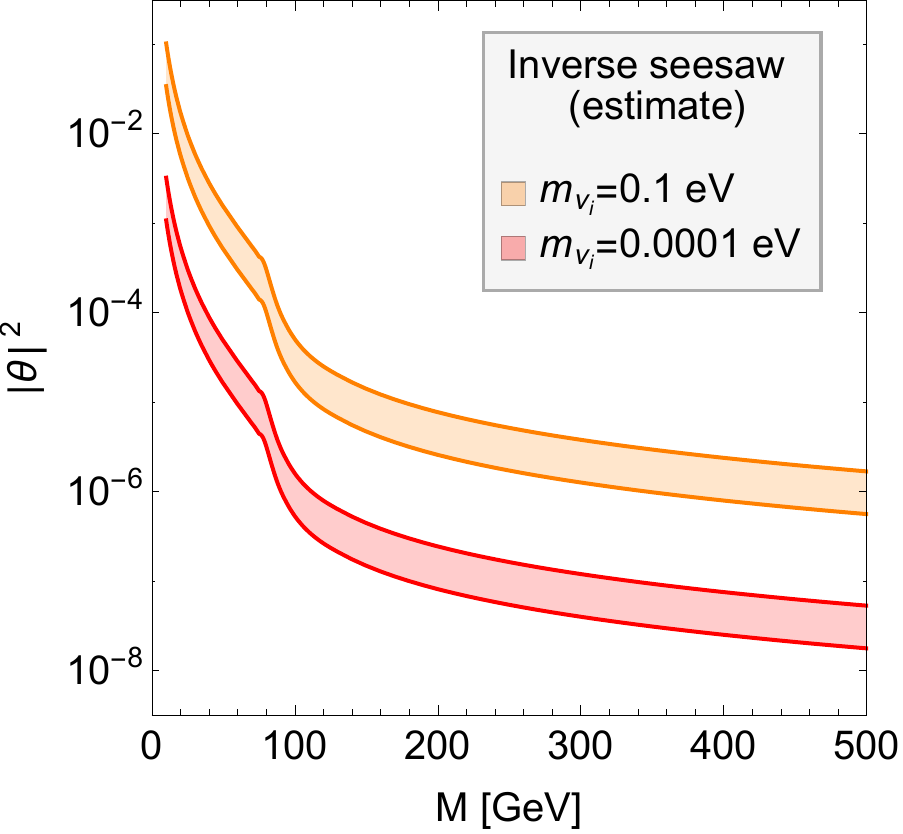}
\caption{$R_{\ell\ell}$ contours for the low scale minimal linear seesaw scenario (using the formula from Eq.~(\ref{eq:Rll}) and predictions for $\Delta M$ from Eqs.~(\ref{eq:DeltaMlinNO}) and (\ref{eq:DeltaMlinIO})) and the inverse seesaw scenario (with estimates for $\Delta M$ from Eq.~(\ref{eq:MnuInvDelta})). For each of the shown bands, the upper contour corresponds to  $R_{\ell\ell} =0.1$ and the lower contour corresponds to  $R_{\ell\ell} =0.9$. In the upper plot we show the case of the minimal linear seesaw with bands for NO and IO, and the lower plot shows the inverse seesaw case with the estimates for $m_{\nu_{i}}= 0.1$ eV and $m_{\nu_{i}}= 10^{-4}$ eV.
Within the bands and below, sizeable LNV due to heavy neutrino-antineutrino oscillations can occur. 
}
\label{fig:Rinfinity}
\end{figure}

\paragraph{Relevance of LNV for collider searches:}
Due to the approximate ``lepton number''-like symmetry it is generally expected that the LNV effects are strongly suppressed. We now quantify under which conditions this statement holds when heavy neutrino-antineutrino oscillations are taken into account, even when the oscillations cannot be resolved experimentally, using the estimates for the mass splittings from Eqs.~(\ref{eq:DeltaMlinNO}), (\ref{eq:DeltaMlinIO}) and Eq.~(\ref{eq:MnuInvDelta}). 
For this purpose, we consider the ratio $R_{\ell\ell} := R_{\ell\ell}(0,\infty)$ (cf.\ \cite{Anamiati:2016uxp}) which can be expressed as 
\begin{equation}\label{eq:Rll}
R_{\ell\ell} = \frac{\Delta M^2}{2 \Gamma^2 + \Delta M^2}\:.
\end{equation}
We calculate the heavy neutrino decay rate $\Gamma$ in our benchmark model using WHIZARD \cite{Kilian:2007gr,Moretti:2001zz}. 
We show, as examples, the range $0.1 \le  R_{\ell\ell}  \le 0.9$ for the low scale minimal linear seesaw scenario (using the formula from Eq.~(\ref{eq:Rll}) and predictions for $\Delta M$ from Eqs.~(\ref{eq:DeltaMlinNO}) and (\ref{eq:DeltaMlinIO})) and the inverse seesaw scenario (with estimates for $\Delta M$ from Eq.~(\ref{eq:MnuInvDelta})) in figure \ref{fig:Rinfinity}. From the figure it can be seen for which parameters $M$ and $|\theta|^2$ sizeable LNV due to heavy neutrino-antineutrino oscillations can occur.

\paragraph{Conclusions:}

We have discussed the possibility to resolve heavy neutrino-antineutrino oscillations at collider experiments, considering low scale seesaw models with a protective ``lepton number''-like symmetry.
We showed that the minimal linear and inverse low scale seesaw scenarios naturally feature almost degenerate heavy neutrino masses which can give rise to oscillation lengths in a (potentially) observable range. 
Especially for the minimal linear seesaw model (with two sterile neutrinos), the oscillation time in the proper frame is fixed by the measured light neutrinos' mass differences (cf.\ Eqs.~(\ref{eq:OscLNO}) and (\ref{eq:OscLIO})) and is thus a prediction that is different for the two orderings.

We focused on signatures of the heavy neutrino-antineutrino oscillations via the processes $pp \to \ell \ell j j $, which result in an oscillating pattern between LNV and LNC dilepton events, i.e. an oscillatory distribution of same-sign and opposite-sign dilepton events. 
We have argued that for the example case of the minimal linear seesaw model with inverse neutrino mass ordering, the predicted oscillation length can lead to observable oscillation signatures at the HL phase of the LHCb and/or at future colliders.
We note that also various other processes can be used to probe heavy neutrino-antineutrino oscillations. Examples are the trilepton final state at the LHC \cite{Dib:2016wge,Dib:2017vux} and basically all final states that feature an unambiguous signal for LNV (for a summary see \cite{Antusch:2016ejd}).
Our general result, that heavy neutrino-antineutrino oscillations can be resolvable at collider experiments, also applies to more general models with at least one pair of heavy neutrinos with sufficiently small mass splitting.

When the heavy neutrino-antineutrino oscillations are fast, such that they are not resolvable experimentally, they can nevertheless give rise to LNV events at colliders. For heavy neutrino masses below the $W$ boson mass, where the HL-LHC and all planned future collider experiments are expected to have best sensitivity (cf.\ \cite{Antusch:2016ejd}), parameter regions where LNV signals are relevant can be reached. For masses above the $W$ boson mass, only for the inverse seesaw estimate we expect that LNV is potentially observable at the currently planned future colliders, while for the minimal linear seesaw models it is safe to assume LNC. 

In summary, we found that heavy neutrino-antineutrino oscillations could be resolvable at colliders. Observing this novel phenomenon would open up a fascinating way to probe sterile neutrino properties and to get insight into the mechanism of neutrino mass generation.

\paragraph{Acknowledgements:} This work has been supported by the Swiss National Science Foundation and it has received funding from the European Unions Horizon 2020 research and innovation programme under the Marie Sklodowska-Curie grant agreement No 674896 (Elusives).
O.F.\ acknowledges support from the ``Fund for promoting young academic talent'' from the University of Basel under the internal reference number DPA2354.

\bibliographystyle{unsrt}

\end{document}